\documentclass[notitlepage
]{revtex4-1}

\usepackage[pdftex]{color,graphicx}
\usepackage{graphicx}
\usepackage[version=3]{mhchem}
\usepackage{mathtools}
\usepackage{setspace}
\usepackage{dcolumn}
\usepackage{bm}

\usepackage{float}
\usepackage{array}
\usepackage{sidecap}
\usepackage{epstopdf}
\usepackage{color}

\begin{document}

\title{Sensitivity Characterisation of a Parametric Transducer for Gravitational Wave Detection Through Optical Spring Effect} %Title of paper

\author{N. C. Carvalho}
\email[]{nataliaccar@gmail.com}
\affiliation{School of Physics, The University of Western Australia, Crawley, 6009, Australia}
\affiliation{ARC Centre of Excellence for Engineered Quantum Systems (EQuS), 35 Stirling Hwy, 6009, Crawley, Australia }
\author{O. D. Aguiar}
\affiliation{Astrophysics Division, National Institute for Space Research – INPE, Av. dos Astronautas 1758, São José dos Campos, Brazil}
\author{J. Bourhill}
\affiliation{School of Physics, The University of Western Australia, Crawley, 6009, Australia}
\affiliation{ARC Centre of Excellence for Engineered Quantum Systems (EQuS), 35 Stirling Hwy, 6009, Crawley, Australia }
\author{M. E. Tobar}
\affiliation{School of Physics, The University of Western Australia, Crawley, 6009, Australia}
\affiliation{ARC Centre of Excellence for Engineered Quantum Systems (EQuS), 35 Stirling Hwy, 6009, Crawley, Australia }

\date{\today}

\begin{abstract}

We present the characterisation of the most recent parametric transducers designed to enhance the Mario Schenberg Gravitational Wave Detector sensitivity. The transducer is composed of a microwave re{-}entrant cavity that attaches to the gravitational wave antenna via a rigid spring. It functions as a three-mode mass-spring system; motion of the spherical antenna couples to a 50 $\mu m$ thick membrane, which converts its mechanical motion into a frequency shift of the cavity resonance. Through the optical spring effect, the microwave transducer frequency-displacement sensitivity was measured to be 726 $MHz/\mu$m at 4 K. The spherical antenna detection sensitivity is determined analytically using the transducer amplification gain and equivalent displacement noise in the test setup, which are 5.5 $\times$ 10$^{11} V/m$  and $1.8 \times 10^{-19} m\sqrt{Hz}^{-1}$, respectively.
\end{abstract}


\maketitle %\maketitle must follow title, authors, abstract and \pacs

\section{Introduction}

Due to the groundbreaking observation of gravitational waves on September 14, 2015, physicists now have access to a brand new probe of the early universe; allowing previously unobservable astronomical events to be monitored and characterised. The proof of gravitational wave{'}s existence has renewed motivation for development of alternative detectors with sensitivities that rival LIGO. 

Although the LIGO detector \cite{abbott2016} has confirmed the interferometric detection system as a successful method to investigate the cosmos through gravitational radiation, large resonant-mass spherical detectors \cite{aguiar2012} still figure as a feasible and resourceful tool, offering a much lower cost model for direct observation. Moreover, these detectors have the advantage of being an isotropic and multi-mode antenna, capable of monitoring an incoming signal from all directions and determining direction, phase, and polarisation of the gravitational waves with a single detector. Such characteristics come from the optimum transducers distribution over the spherical surface, which localize six of them in a truncated icosahedron configuration \cite{merkowitz1995}. Hence, the antenna becomes onmidirectional and has different transducers monitoring different modes, allowing a single antenna to act as a number of detectors operating at the same location. For these reasons, it is expected that the combination of the spherical detectors with the current realisations will net an invaluable contribution to the field \cite{aguiar2006, costa2006}.

Aside from achieving the ultimate goal of gravitational wave detection, this field of research has proved invaluable over its lifetime through consistently outputting state of the art spin-off technologies with diverse applications, such as low noise oscillators \cite{ivanov2006}, frequency standards technology \cite{locke2008}, as well as laying much of the foundation work for the modern field of cavity optomechanics \cite{Cuthbertson96, tobar1997}. The latter arose due to resonant-mass gravitational wave detectors employing parametric transducers \cite{tobar1993, linthorne1992, tsubono1986}, such as those used in the Mario Schenberg detector and the Australian antenna Niobe \cite{tobar2000}. This type of transducer is used to transfer energy from acoustic to optical (or microwave) frequencies by changing a resonant cavity length through the mechanical vibration of one of its boundaries. In gravitational wave detectors, these devices convert the low amplitude vibration caused by an incident gravitational wave into a measurable electrical signal, making possible the detection and characterisation of the precursor astronomical event \cite{aasi2013, abbott2016, de2015}. Various resonant antenna groups, such as Allegro \cite{mauceli1996}, Auriga \cite{zendri2002}, Explorer \cite{astone2002}, Minigrail \cite{gottardi2007} and Nautilus \cite{astone1997}, used different types of transducers, but also contributed greatly to derivative technologies in gravitational wave research.

Other realisations of these optomechanical devices, which can range in size from mesoscopic to macroscopic regimes, have demonstrated extremely impressive results, such as ground state cooling \cite{teufel2011, abbott2009, schliesser2006}, the observation of non-classical effects in macroscopic objects \cite{marshall2003, schliesser2010, bourhill2015}, as well as a profusion of possibilities for additional applications in innovative technologies \cite{cleland1998, naik2009, stannigel2010, yang2006, peng2006, rabl2010}. This field currently presents itself as the leading pathway for quantum mechanical analysis of displacements and hence precision sensing \cite{schliesser2008, aspelmeyer2014, teufel2011}, and so there is a great motivation to improve their performance through obtaining remarkably high mechanical and electrical quality factors and highly efficient thermal and phase noise suppression.

Here, we report the frequency-displacement sensitivity characterisation of the latest transducer design of the Mario Schenberg Gravitational Wave Detector. Also, we discuss and estimate the detection sensitivity with the use of a set of similar transducers. These devices are to be attached to a 1150 kg CuAl (6\%) sphere constructed to detect events in the 3.1 - 3.3 kHz frequency bandwidth, such as core collapse in supernova events, coalescence of neutron stars, black hole systems and more \cite{aguiar2012}.
 
\begin{figure}
\begin{center}
	\includegraphics[height=4.5cm]{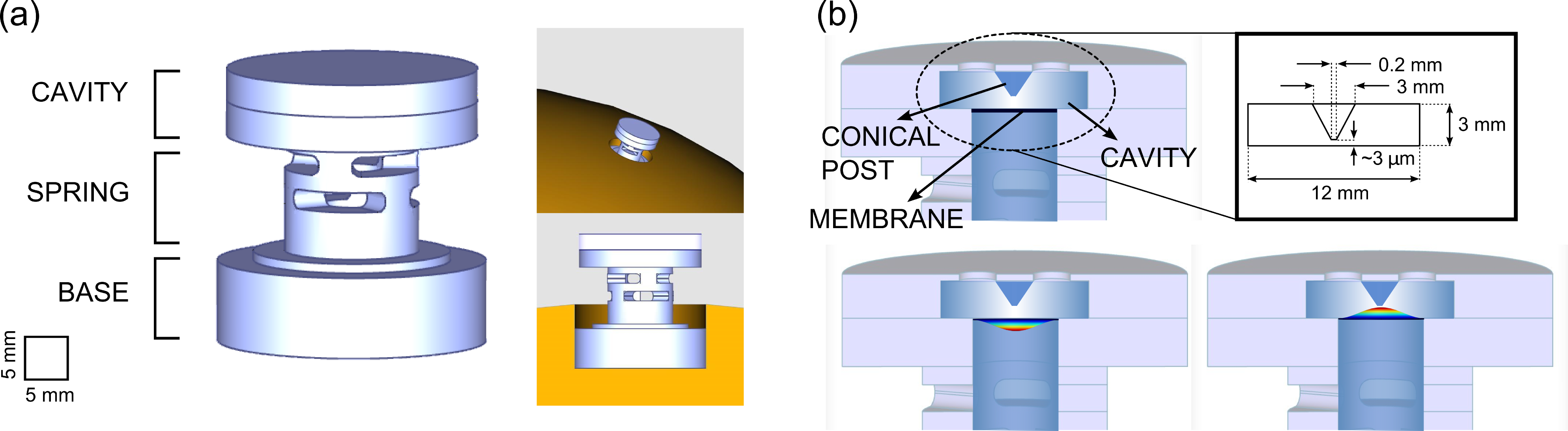}
	\caption{Schenberg's transducer design: (a) Three-dimensional representation. Insets: fixture to the spherical antenna scheme; (b) Top: Cross-section of the transducer's cavity and its dimensions; Bottom: Exaggerated membrane motion.\label{fig1}}
\end{center}
\end{figure}

The Schenberg's transducer, made of niobium and schematically represented in Fig. \ref{fig1}(a), comprises of a base, spring, and a microwave cavity. As the detector is designed to operate at cryogenic temperatures, the base is placed inside a hole on the sphere surface; using the differential thermal contraction between the sphere and transducer material to secure the transducer \cite{de2015} (as in Fig. \ref{fig1}(a)) inset). The whole set operates as a three-mode mass-spring system, which has been shown to improve the detector bandwidth and performance \cite{Richard1984, Bassan1988, Marchese1994, tobar1995}. In this case, the first mode is the spherical antenna, the second is the mechanical structure of the transducer, and the third is a 50 $\mu$m thick niobium membrane, whose function is to close one end of the cylindrical cavity.

The cavity has a reentrant design (often referred as Klystron \cite{fujisawa1958, de2013}). When the transducer experiences a mechanical vibration at a specific frequency, the membrane{'}s drum mode is excited, and the gap spacing between it and the top of the conical post oscillates, modulating the intra-cavity field. A diagram with the cavity dimensions and the membrane motion are shown in Fig. \ref{fig1}(b). Also, a more detailed description of the transducer design can be found in \cite{de2015}. 

\section{Experiment}

\begin{figure}
\begin{center}
	\includegraphics[height=3.8cm]{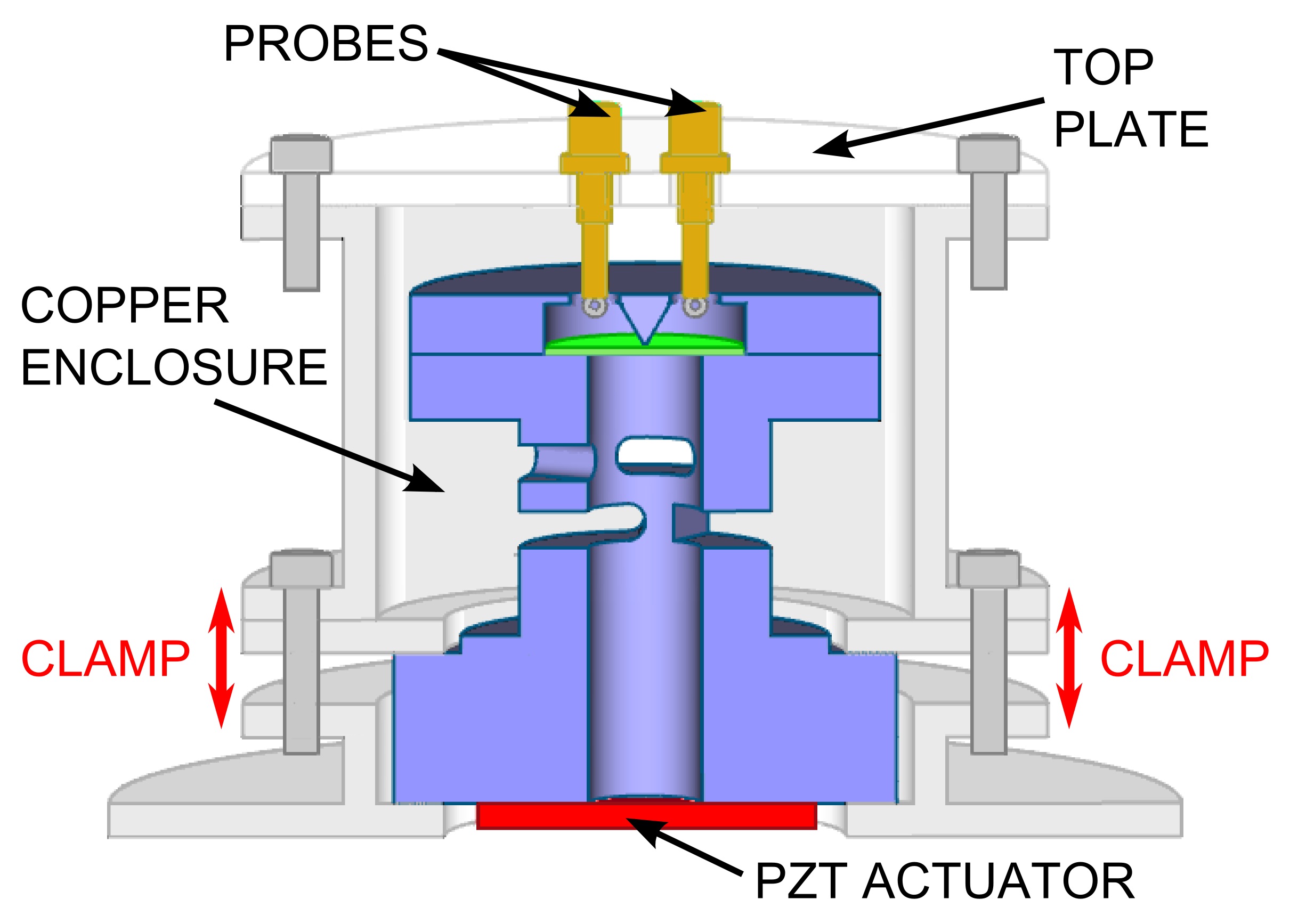}
	\caption{Cross-section of the test enclosure used to characterise the Schenberg's transducer (PZT: piezoelectric actuator).\label{fig2}}
\end{center}
\end{figure}

For practical reasons, a test setup was developed to characterise the transducer{'}s frequency-displacement sensitivity without the need to cool the entire detector. A cross-section of the temporary test enclosure is shown in Fig. \ref{fig2}. It is mostly copper except for the top plate, which is polytetrafluoroethylene to avoid the spurious electromagnetic modes that would be generated in a closed metallic enclosure. 

The objectives of the enclosure were: i) to attach the transducer to the refrigerator cold finger and ii) to provide a contact-less way to couple two magnetic probes to the microwave cavity field. Also, to reproduce the conditions of attachment to the sphere, the transducer was only clamped by its base, and to excite the mechanical modes of the system, a piezoelectric actuator was fixed to the bottom of the transducer's base. 

The transducer was cooled in a Blue Fors dilution refrigerator system. However, the system was operating at 4 K, i.e., only the Pulse Tube cryocooler was running and the dilution refrigerator cycle was off. A Rox$^{TM}$ temperature sensor located at the same refrigerator's flange as the transducer was used to monitor it.

Altogether, the detector can accommodate nine transducers. The transducer chosen to be sampled was designed to operate at a mechanical frequency of approximately 6.5 kHz. Although the detector target frequency is 3.2 kHz and most of the transducers were built to operate at this frequency, a subset were chosen to monitor the proposed mono-polar excitation predicted by alternative gravity theories \cite{forward1971, bianchi1996} at 6.5 kHz. 

The fact that the transducers operate at different frequencies does not restrict the achieved results from being extended to the whole set, as the focus of this investigation is the microwave cavity displacement sensitivity, which only depends on the cavity geometry. The nine microwave cavities supported by the springs have equal dimensions, all of them built to resonate around 10 GHz when the system reaches liquid helium temperature. 

\subsection{Mechanical resonances and quality factors}

\begin{figure}
\begin{center}
	\includegraphics[height=7cm]{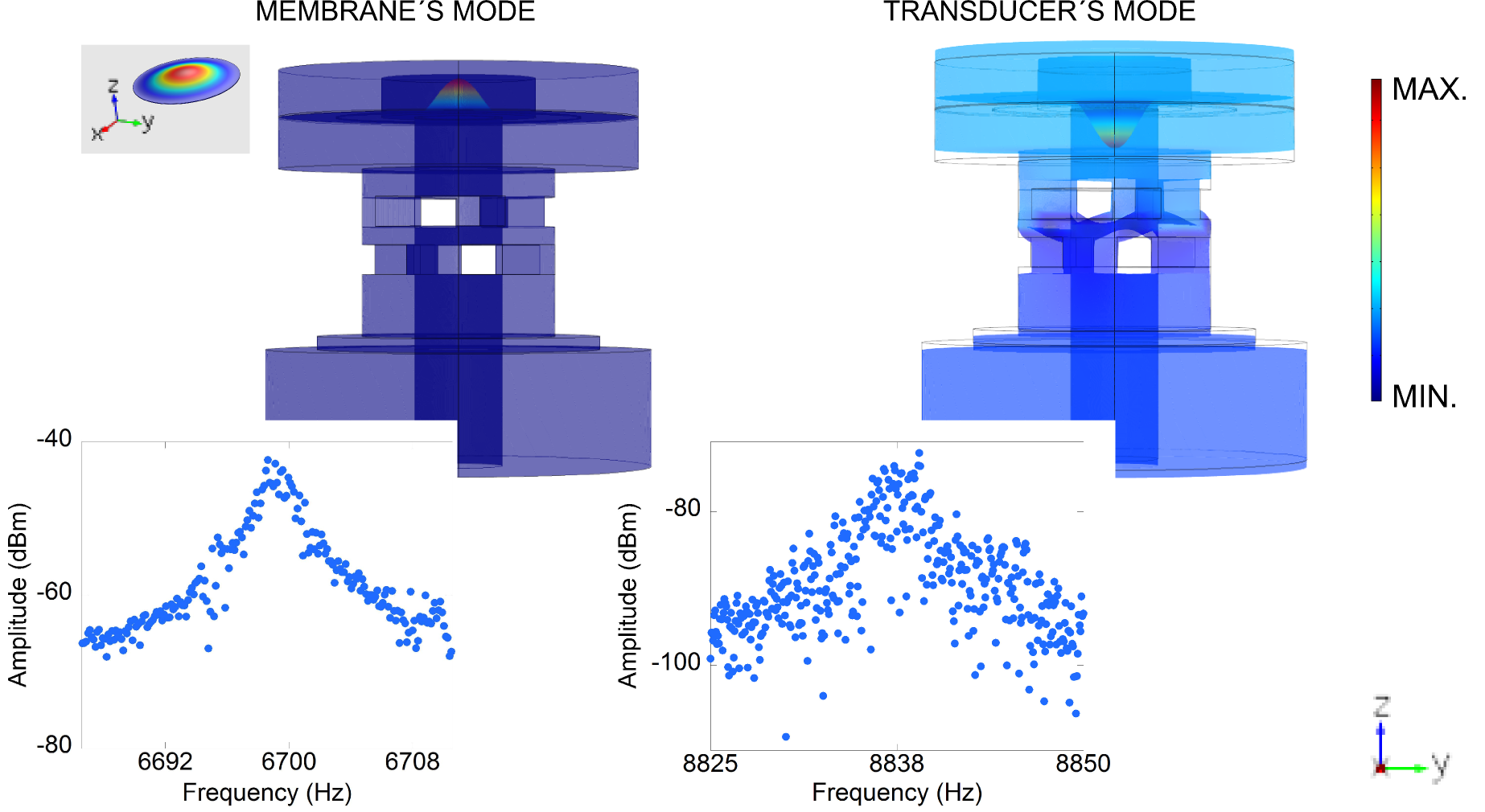}
	\caption{Displacement density plots of the mechanical modes obtained from FEM simulation. Insets: Simulated membrane drum mode and measured membrane and transducer modes. Colour bar: mechanical displacement field. \label{fig3}}
\end{center}
\end{figure}

With the aid of a Vector Network Analyser connected in transmission to the cavity, the microwave re-entrant mode was found at 9.539 GHz with a loaded Q-factor of 1677 at 4 K. In order to verify the mechanical frequencies of the membrane drum mode and the transducer longitudinal mode, shown in Fig. \ref{fig3}, a digital synthesiser was connected to the input of the microwave cavity to pump its resonance while an AC source drove the piezoelectric actuator with a chirp signal between 1 and 10 kHz. 

\begin{figure}
\begin{center}
	\includegraphics[height=5cm]{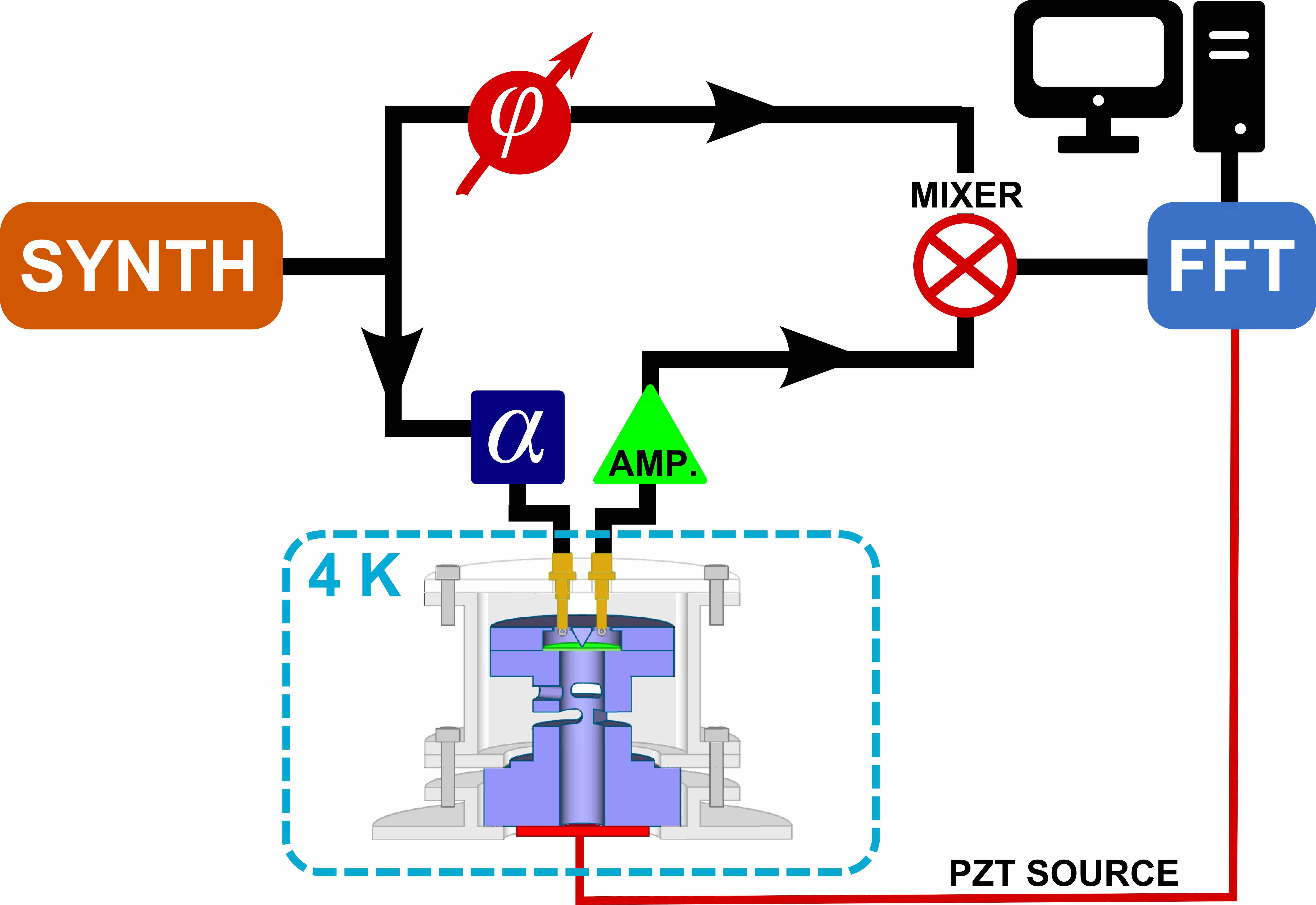}
	\caption{Schematic diagram of the frequency-discriminator used to characterise the transducer and membrane mechanical resonances and determine the transducer's frequency-displacement sensitivity (Synth: RF synthesiser; $\phi$: mechanical phase-shifter; Amp.: microwave amplifier; $\alpha$: microwave attenuator; PZT: piezoelectric). \label{fig4}}
\end{center}
\end{figure}

The cavity output microwave signal was sent to a frequency discriminator represented by the diagram in Fig. \ref{fig4}. Here, the discriminator functions as a homodyne detector, converting frequency fluctuations caused by the mechanical motion into voltage fluctuations. This allows the mechanical modes to be read out on a Fast Fourier Transform Spectrum Analyser (FFT). 

Two prominent mechanical modes were found at 6.69 kHz and 8.84 kHz, with Q-factors ($Q_m$) of 4500 and 2200, respectively. Room temperature measurements \cite{de2015}, however, had previously shown that the transducer longitudinal mode was correctly tuned to the designed frequency. Thus, to understanding why the 4 K results differ from 6.5 kHz, Finite Element Modelling (FEM) software was employed. Simulation results showed that the mechanical resonance of the transducer's body would be detuned from its designed frequency if the enclosure failed to hold the transducer's base rigidly {--} this was simulated by modelling the transducer with both a fixed and free boundary condition. The model confirms that a {``}free{''} transducer resonates at higher frequencies as observed in the experiment. 

Therefore, the transducer longitudinal mode was not perfectly clamped and could not be precisely modelled. Despite this, the membrane experimental boundary conditions were totally reproducible, leading to the conclusion that the membrane's drum mode frequency corresponds to the 6.690 kHz observed mode, measured with a discrepancy of only 3\% compared to the estimated value. Such a result is within the tolerance and therefore does not affect the transducers displacement sensitivity significantly.

\subsection{Transducer frequency-displacement sensitivity}

The transducer frequency-displacement sensitivity can be determined by observing the mechanical frequency shift when the microwave pump is blue and red detuned from the cavity resonance via the optical spring effect. The resulting frequency shift of the mechanical mode originates from radiation back-action force. To derive this relation, one has to solve the linearised equations of motion for the light and mechanics in the frequency space; this will allow the definition of the modified mechanical susceptibility that accounts for the optomechanical interaction. 

By taking the real part of the adjusted susceptibility, the frequency-dependent mechanical frequency shift caused by the spring effect is obtained \cite{aspelmeyer2014}:

\begin{equation}
\footnotesize
	\delta\Omega(\omega) = g_0^2 n_{cav}\left(\frac{\Delta-\Omega}{\kappa^2/4 + (\Delta-\Omega)^2}-\frac{\Delta+\Omega}{\kappa^2/4 + (\Delta-\Omega)^2}\right),
	\label{Eq1}
\end{equation}

\noindent where $\Delta$ is the difference between the pump frequency, $\omega$,  and the cavity resonant frequency, $\omega_0$; $\Omega$ is mechanical frequency and $\kappa$ is the overall cavity intensity decay rate.

The optomechanical single-photon coupling strength, $g_0$, is given by:

\begin{equation}
	g_0 = -2\pi x_{zpf} \frac{df}{dx}
	\label{Eq2}
\end{equation}

Here, $\frac{df}{dx}$ is the cavity frequency-displacement sensitivity, where $f = 2\pi\omega$,

\begin{equation}
	x_{zpf} = \sqrt{\frac{\hbar}{2 m_{\textit{eff}} \Omega}},
	\label{Eq4}
\end{equation}

\noindent is the mechanical zero-point fluctuation amplitude and $m_{\textit{eff}}$  is the effective mass of the mechanical mode.

The cavity photon number, $n_{cav}$, is

\begin{equation}
	n_{cav} = \frac{P_{in}}{\hbar\omega_0} \left(\frac{\kappa}{(\kappa/2)^2 + \Delta^2}\right) \text{with }  \kappa = \kappa_0(1 + \beta_1 +\beta_2),
	\label{Eq3}
\end{equation}

\noindent where $P_{in}$ is the cavity input power, $\beta_1$ and $\beta_2$ are the cavity input and output couplings, respectively, and $\kappa_0$ represents the cavity intrinsic losses.

\begin{figure}
\begin{center}
	\includegraphics[height=5cm]{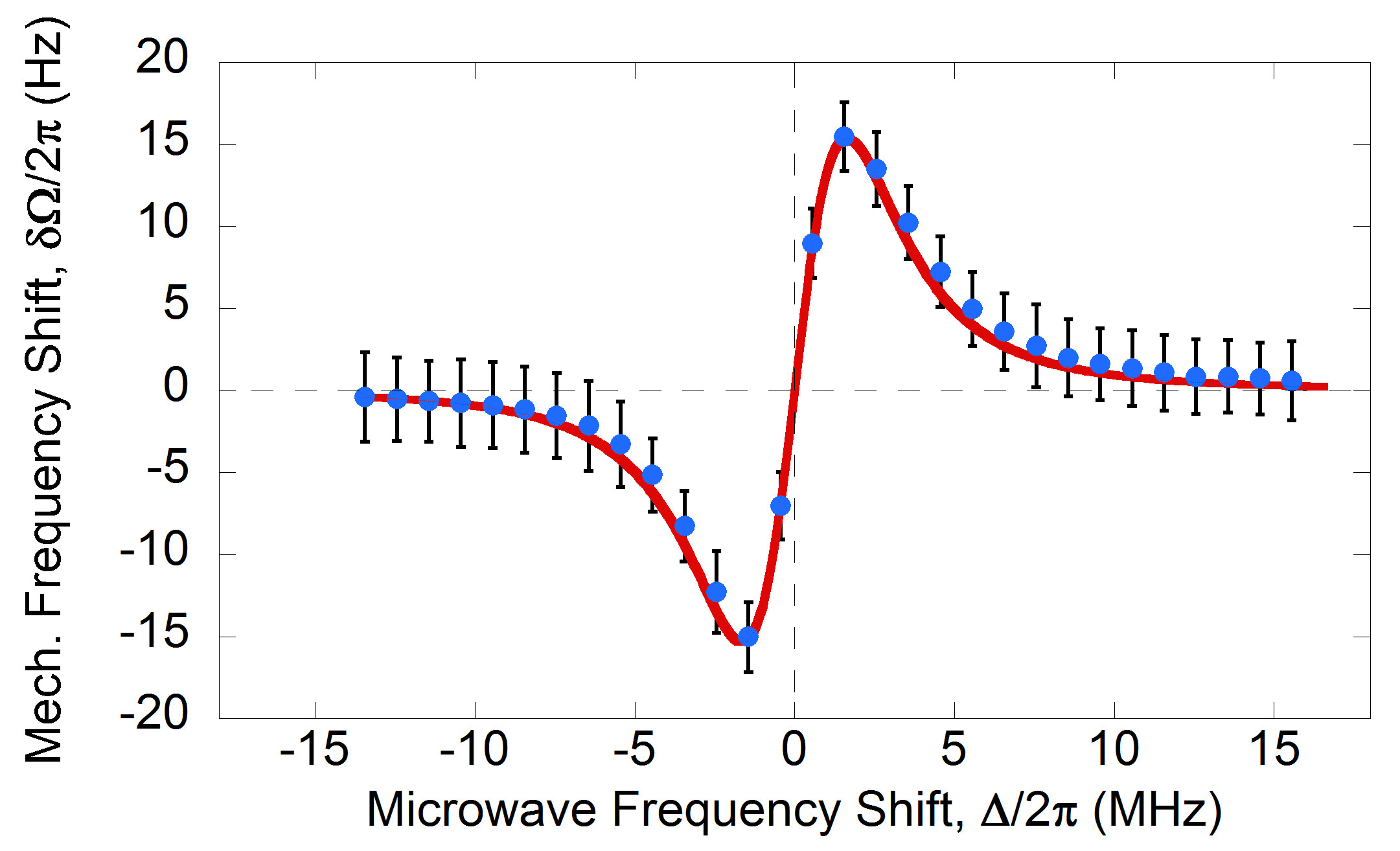}
	\caption{Effect of the microwave detuning on the mechanical frequency (signature of the optical spring effect). Solid line: Nonlinear fit used to determine the frequency displacement sensitivity. Points: measured data. \label{fig5}}
\end{center}
\end{figure}

The value of $\frac{df}{dx}$ was calculated from (\ref{Eq1}) using the membrane mode. This was achieved using the set up shown in Fig. \ref{fig4}; frequency response traces of the mechanical mode were recorded while the pump frequency was detuned from the cavity resonance. Each trace was fitted by a Lorentzian function, which gave a very good estimate of the mechanical frequencies. The results are shown in Fig. \ref{fig5}.

To fit the data points with (\ref{Eq1}) it was first necessary to determine $x_{zpf}$ and $n_{cav}$. To calculate the former, the effective mass of the membrane drum mode must be obtained from the FEM, which gave $m_{\textit{eff}}= 7.061$ mg. Then, using the mechanical frequency of this mode, the $x_{zpf}$ was found to be $1.3329 \times 10^{-17}$ m. The average number of photons circulating into the cavity is a function of the cavity detuning. The constant terms, however, can be calculated from the experimental parameters: cavity input power, which was $P_{in}=6.8$ mW, and the input and output couplings, $\beta_1=0.006$ and $\beta_2=0.0025$, respectively.   

Finally, using the best fit of the data, it was found that $\frac{df}{dx} = 726 \pm 221$ MHz$/\mu$m, which is the transducer frequency-displacement sensitivity. The errors originate from measurement uncertainties in the couplings, input power and, loaded electrical Q-factor. The fitting is represented by the solid line in Fig. \ref{fig5} and shows a standard error of 6 MHz/$\mu$m.

According to the FEM for the electromagnetic field into the cavity, the frequency-displacement sensitivity expected for the reentrant mode was 770 MHz/$\mu$m, calculated by changing the cavity gap spacing in a narrow range around the designed gap and recording the respective re-entrant mode frequency for each gap size. The experimental result deviates less than 6\% from the simulated design. 

\subsection{Conversion efficiency of frequency to voltage}
	
To calculate the detector sensitivity using the above transducer design the conversion efficiency of frequency to voltage,  $\frac{du}{df}$, must be known. This parameter is dependent on the electrical Q-factor and cavity input power. Although the results cannot be generalized as the detector experimental conditions are not being completely reproduced, it will offer a good way to estimate the overall sensitivity that the detector can reach.

\begin{figure}
\begin{center}
	\includegraphics[height=6cm]{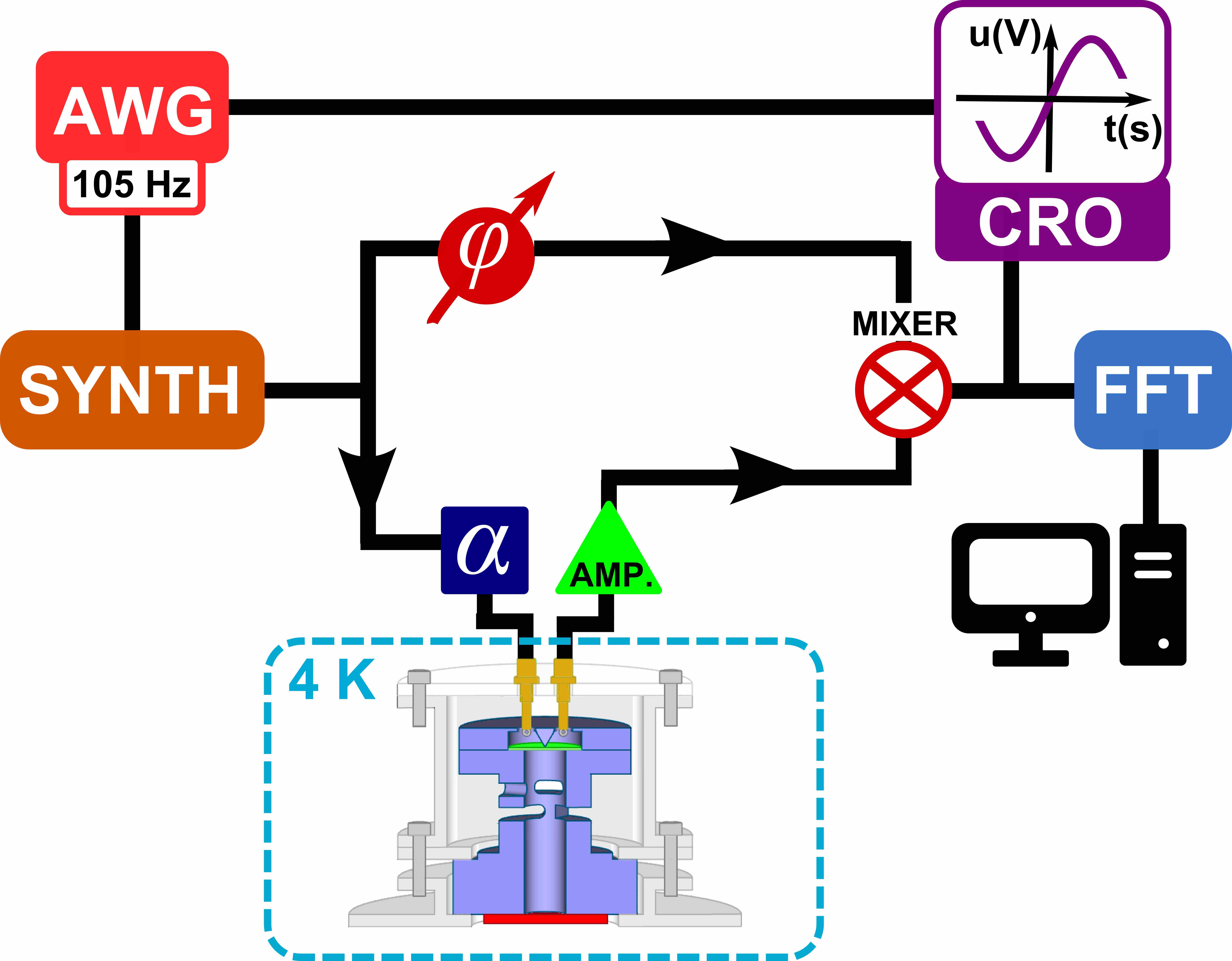}
	\caption{Schematic diagram of the modified frequency-discriminator used to determine the transducer's conversion efficiency of frequency to voltage. (Synth: RF synthesizer; $\phi$: mechanical phase-shifter; Amp.: microwave amplifier; $\alpha$: microwave attenuator; AWG: Arbitrary wave generator; CRO: cathode-ray oscilloscope). \label{fig6}}
\end{center}
\end{figure}

In this paper, $\frac{du}{df}$ was measured using the apparatus shown in Fig. \ref{fig6}. Here the cavity resonance was excited by the synthesizer and artificially modulated at 105 Hz ($t_{mod}$ = 9.524 ms) at a rate, $r_{mod}$, equal to 16 MHz. As a result, an error curve could be read out in the oscilloscope, giving the voltage variation with time,  $\frac{du}{dt}$ , at the output of the frequency discriminator.

Strictly speaking, the conversion efficiency of frequency to voltage is dependent on the modulation frequency, $f_{mod}$, however, if $f_{mod} << \kappa_0/2$, which is true as for the artificial modulation as for the membrane drum mode, it will be approximately constant over the frequency range under consideration \cite{ivanov2006}. In this case, $\frac{du}{df}$ can be given by:

\begin{equation}
	\frac{du}{df} = \frac{du}{dt} \frac{t_{mod}}{r_{mod}}. 
	\label{Eq5}
\end{equation}

\begin{figure}
\begin{center}
	\includegraphics[height=8cm]{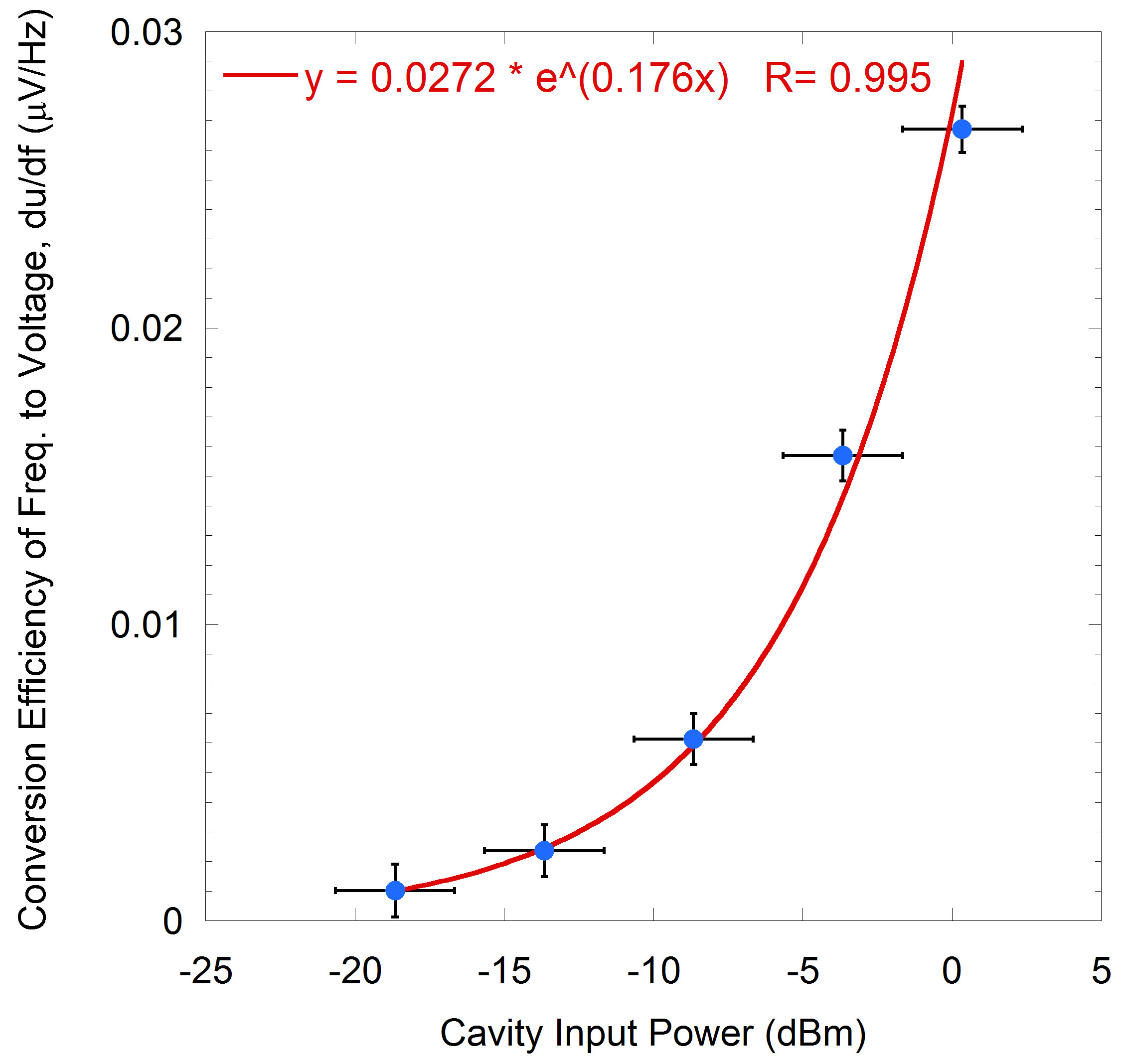}
	\caption{Measured conversion efficiency of frequency to voltage. Solid line: Exponential fit (equation shown on left corner). \label{fig7}}
\end{center}
\end{figure}

Using (\ref{Eq5}),  $\frac{du}{df}$ was calculated for different values of cavity input power as shown by Fig. \ref{fig7}.  Then, by fitting the data, the conversion efficiency of frequency to voltage at 6.8 mW (8.3 dBm) was found to be 118 $\pm$ 1 mV/MHz, where the uncertainty is given by the standard error.

\subsection{Transducer amplification gain and equivalent displacement noise}

The transducer amplification gain, $\Gamma$, is obtained by multiplying the conversion efficiency of displacement to voltage, $\frac{du}{dx}$, by the intrinsic amplitude gain of the transducer-sphere system, $\gamma$. The first is given by

\begin{equation}
	\frac{du}{dx} = \frac{df}{dx} \frac{du}{df} 
	\label{Eq6}
\end{equation}

\noindent and the second is

\begin{equation}
	\gamma = \sqrt{\frac{M}{m_{eff}}},
	\label{Eq7}
\end{equation}

\noindent where M is the sphere modal mass, equal to 287.5 kg.

Solving (\ref{Eq6}) and (\ref{Eq7}), $\gamma$ and $\frac{du}{dx}$ are respectively, 6381 and $85.7 \times 10^6$ V/m. Hence, the amplification gain of the current transducer design, subjected to the test experimental conditions, is approximately $5.5 \times 10^{11}$ V/m.

The equivalent displacement noise of the transducer is used to calculate how sensitively the detector can distinguish a gravitational wave signal above white noise. One has to divide the power spectrum density (PSD) background level by $\Gamma$. Here, the PSD background level, seen in the FFT when measuring the mechanical frequencies, is of the order of $1 \times 10^{-7}$ V/$\sqrt{Hz}$. Therefore, the equivalent displacement noise is approximately $1.8 \times 10^{-19}$ m/$\sqrt{Hz}$. 

In the test setup, the noise level is considerably higher than what is expected in the detector; this is because the transducer was not held isolated from the laboratory vibrations and the electronic apparatus was not optimised for low noise operation. For this reason, it was not possible to observe the effect of the Langevin forces due to the Brownian motion. Still, supposing the detector were to operate under the current test conditions, it is possible to estimate its sensitivity if we divide the equivalent displacement noise by the sphere radius, 0.325 m. Then, this would result in a detection sensitivity of $5.5 \times 10^{-19}\sqrt{Hz^{-1}}$.

The detection system aims for a sensitivity of the order of $10^{-22}\sqrt{Hz^{-1}}$, which means that increasing $\frac{du}{df}$ one order of magnitude and decreasing the background level one order of magnitude would result in an operation very close to the optimum scenario. Therefore, we believe it is possible to reach the design sensitivity in the near future with the aid of a few improvements: the Mario Schenberg antenna can be cooled down to 0.1 K, decreasing its Brownian noise; its vibration isolation with proper state-of-the-art operation can be improved at least one order of magnitude; electric Q-factors of $3 \times 10^5$, which were already achieved for some transducers, should be guaranteed for the whole set of transducers; and the transducer mechanical Q-factors may be increased one or two orders of magnitude by annealing.

\subsection{Optomechanical performance}

It is useful to compare the Schenberg{'}s transducer to other optomechanical systems. Two parameters in particular are of critical importance in this respect, namely the degree of decoupling to the thermal environment, given by $Q_m\times\Omega/2\pi$ (or $Q\times f$ product), and the vacuum optomechanical coupling strength, $g_0=\frac{df}{dx}x_{zpf}$. Specifically, if the former fulfils the condition $\hbar Q_m\times \Omega/2\pi > {k_B T}$, which at mK temperatures corresponds to the condition $Q_m\times \Omega/2\pi>83$ GHz, one may neglect thermal decoherence over one mechanical period, whilst the latter refers to the cavity frequency shift induced by a mechanical zero-point displacement; i.e. caused by the interaction of a single phonon with a single photon within the particular system. 

For the described transducer, $g_0=9.7$ mHz and $Q\times\Omega/2\pi=3\times10^7$ Hz. Although these values are consistent with other published cavity optomechanics experiments \cite{aspelmeyer2014}, there is the potential for further optimisation. $g_0$ could be increased by reducing the mass of the membrane, whilst obtaining a higher mechanical quality factor for the resonator would push the transducer further towards the quantum regime.

\section{Conclusion}

We reported the characterisation of the microwave parametric transducer developed to be used in the Mario Schenberg Gravitational Wave Detector. Through the investigation of the optical spring effect, the frequency-displacement sensitivity of the transducer was measured to be 726 MHz/$\mu$m. The experimental result presented a standard error of only 6 MHz/$\mu$m and deviated less than 6\% from the simulated sensitivity.

Furthermore, we also examined the Schenberg's detection sensitivity by measuring the transducer amplification gain and equivalent displacement noise in the test setup, obtaining $5.5 \times 10^{11}$ V/m and $1.8 \times 10^{-19}$ m/ $\sqrt{Hz}$, respectively. Based on this result, we estimate that the detector can reach its designed sensitivity of 10$^{-22} \sqrt{Hz^{-1}}$ around 3.2 kHz and 6.5 kHz with relatively small improvements. 

The mechanical $Q \times f$ product and the vacuum optomechanical coupling strength were also calculated and the values of $3 \times 10^7$ Hz and 9.7 mHz, respectively, show that this transducer design still offers room for further optimisation and, therefore, higher detection sensitivities.

\begin{acknowledgments}

We thank to Dr. Leandro A. N. de Paula for bringing the transducer from Brazil to Australia and our research supporters: the Australian Research Council Grant No. CE110001013, the CNPq and  the FAPESP under grants No. 1998/13468-9 and 2006 /56041-3.

\end{acknowledgments}

%\bibliography{phdref2}


\end{document}